\documentclass[english,aps,prb,twocolumn,superscriptaddress,showpacs,showkeys]{revtex4-1}
\usepackage[T1]{fontenc}
\usepackage{color}
\usepackage{babel}
\usepackage{booktabs}
\usepackage{units}
\usepackage{textcomp}
\usepackage{amstext}
\usepackage{graphicx}
\usepackage[unicode=true,
 bookmarks=true,bookmarksnumbered=false,bookmarksopen=false,
 breaklinks=false,pdfborder={0 0 1},backref=false,colorlinks=true]
 {hyperref}
\hypersetup{pdftitle={Uranium ferromagnet with negligible magnetocrystalline anisotropy- U4Ru7Ge6},
 pdfauthor={M. Valiska, M. Divis, V. Sechovsky},
 pdfkeywords={Keywords: Itinerant 5f-electron ferromagnetism, low anisotropy, magnetostriction}}
\usepackage{breakurl}

\makeatletter

\providecommand{\tabularnewline}{\\}

\makeatother

\begin{document}

\title{Uranium ferromagnet with negligible magnetocrystalline anisotropy
- $\mathrm{U_{4}Ru_{7}Ge_{6}}$}

\author{Michal Vali\v{s}ka}
\email{michal.valiska@gmail.com}

\selectlanguage{english}%

\author{Martin Divi\v{s}}

\author{Vladimír Sechovský}

\affiliation{Faculty of Mathematics and Physics, Charles University, DCMP, Ke
Karlovu 5, CZ-12116 Praha 2, Czech Republic}
\begin{abstract}
Strong magnetocrystalline anisotropy is a well-known property of uranium
compounds. The almost isotropic ferromagnetism in $\mathrm{U_{4}Ru_{7}Ge_{6}}$
reported in this paper represents a striking exception. We present
results of magnetization, AC susceptibility, thermal expansion, specific
heat and electrical resistivity measurements performed on a $\mathrm{U_{4}Ru_{7}Ge_{6}}$
single crystal at various temperatures and magnetic fields and discuss
them in conjunction with results of first-principles electronic-structure
calculations. $\mathrm{U_{4}Ru_{7}Ge_{6}}$ behaves as an itinerant
5$f$-electron ferromagnet ($T_{\mathrm{C}}=\unit[10.7]{K}$, $\mu_{\mathrm{S}}=\unit[0.85]{\mu_{\mathrm{B}}/f.u.}$
at $\unit[1.9]{K}$. The ground-state easy-magnetization direction
is along the {[}111{]} axis of the cubic lattice. The anisotropy field
$\mu_{0}H_{\mathrm{a}}$ along the {[}001{]} direction is only of
$\unit[\sim0.3]{T}$, which is at least 3 orders of magnitude smaller
value than in other U ferromagnets. At $T_{r}=\unit[5.9]{K}$ the
easy magnetization direction changes for {[}001{]} which holds at
temperatures up to $T_{\mathrm{C}}$. This transition causing a change
of magnetic symmetry is significantly projected in low-field magnetization,
AC susceptibility and thermal-expansion data whereas only a weak anomaly
is observed at $T_{r}$ in the temperature dependence of specific
heat and electrical resistivity, respectively. The magnetoelastic
interaction induces a rhombohedral (tetragonal) distortion of the
paramagnetic cubic crystal lattice in case of the {[}111{]}({[}001{]})
easy-magnetization direction. The rhombohedral distortion is connected
with two crystallographically inequivalent U sites. The ab initio
calculated ground-state magnetic moment of $\unit[1.01]{\mu_{\mathrm{B}}/f.u.}$
is oriented along {[}111{]}. The two crystallographically inequivalent
U sites are a consequence of spin-orbit coupling of the U 5$f$-electrons.
In the excited state which is only $\unit[0.9]{meV}$ above the ground
state the moment points to the {[}001{]} direction in agreement with
experiment. A scenario of the origin of the very weak magnetic anisotropy
of $\mathrm{U_{4}Ru_{7}Ge_{6}}$ is discussed considering interactions
of the U-ion 5$f$-electron orbitals with the nearest neighbor ions. 
\end{abstract}


\maketitle

\section{Introduction}

Magnetocrystalline anisotropy (MA) is manifested by locking the magnetic
moments in a specific orientation (usually the easy magnetization
direction) with respect to crystal axes. A quantitative measure of
MA, the anisotropy field $H_{\mathrm{a}}$, is the magnetic field
needed to be applied in the hard magnetization direction for reaching
the easy-axis magnetization value. The key prerequisites of MA are
the orbital moment of a magnetic ion, the spin-orbit (s-o) interaction
coupling the orbital and spin moment, and interactions with neighboring
ions\cite{RN34,RN35}. The s-o interaction is a relativistic effect
which becomes stronger in heavier atoms. Consequently MA dominates
magnetism in materials with lanthanide and actinide ions bearing magnetic
moments of the 4$f$- and 5$f$-electrons, respectively. 

The widely accepted scenario of the origin of magnetocrystalline anisotropy
involves the crystal electric field (CEF) interaction, the single-ion
mechanism born in the electrostatic interaction of the anisotropic
crystalline electric field (the potential created at the magnetic
ion site by the electric charge distribution in rest of the crystal)
with the aspherical charge cloud of the magnetic electrons. The electron
orbital adopts the direction that minimizes the CEF interaction energy.
The single-ion anisotropy is most often encountered in compounds with
lanthanides having the well-localized 4$f$-electrons\cite{RN36,RN38}.

Contrary to the 4$f$-orbitals deeply buried in the core electron
density of lanthanide ions the spatially extended uranium 5$f$-electron
wave functions interact with the overlapping 5$f$-orbitals of the
nearest neighbor U ions (5$f$-5$f$ overlap) and with valence electron
orbitals of ligands (5$f$-ligand hybridization\cite{RN39}). Consequently
the 5$f$-electron wave functions loose the atomic character and simultaneously
the U magnetic moments melt down. Despite that the strong spin-orbit
coupling induces a predominant orbital magnetic moment antiparallel
to the spin moment in the spin-polarized 5$f$-electron energy bands.
This effect was first demonstrated for the itinerant 5$f$-electron
magnetism in $\mathrm{UN}$\cite{RN40}. In some cases very small
or eventually zero total magnetic moment of a U ion are observed as
a result of mutual compensation of the antiparallel spin and orbital
component, respectively. The itinerant 5$f$-electron ferromagnet
$\mathrm{UNi_{2}}$\cite{RN41} with the U magnetic moment of few
hundredths $\mu_{\mathrm{B}}$ serves as an excellent example as documented
by results of polarized neutron measurements\cite{RN42} and first
principles electronic structure calculations\cite{RN43}. Despite
the itinerant character of magnetism $\mathrm{UNi_{2}}$ exhibits
very strong magnetocrystalline anisotropy with $\mu_{0}H_{\mathrm{a}}\gg\unit[35]{T}$
at $\unit[4.2]{K}$\cite{RN44}.

The very strong MA seems to be inherent to the uranium magnetism.
The typical values of the MA field of most uranium intermetallic compounds
are of the order of hundreds $\mathrm{T}$\cite{RN20}. The strong
anisotropy is reported also for the cubic U pnictides and chalcogenides\cite{RN26,RN45}. 

The strong interaction of the spatially extended U 5$f$ orbitals
with surrounding ligands in the crystal and participation of 5$f$
electrons in bonding\cite{RN46,RN47} imply an essentially different
mechanism of magnetocrystalline anisotropy based on a two ion (U-U)
interaction. The anisotropy of the bonding and 5$f$-ligand hybridization
assisted by the strong spin-orbit interaction are the key ingredients
of the two-ion anisotropy in 5$f$-electron magnets. 

The systematic occurrence of particular types of anisotropy related
to the layout of the U ions in a crystal lattice suggests that in
materials, in which the U-U co-ordination is clearly defined in the
crystal structure, the easy-magnetization direction is perpendicular
to the nearest U-U links\cite{RN20,RN21}. Cooper and co-workers\cite{RN48,RN49}
have formulated a relatively simple model of the two-ion interaction.
This model, leading to qualitatively realistic results, is based on
Coqblin-Schrieffer approach to the mixing of ionic $f$-states and
conduction-electron states, in which the mixing term of the Hamiltonian
of Anderson type is treated as a perturbation, and the hybridization
interaction is replaced by an effective $f$-electron-band electron
resonant exchange scattering. The theory has been further extended
so that each partially delocalized $f$-electron ion is coupled by
the hybridization to the band electron sea; and these both lead to
a hybridization-mediated anisotropic two-ion interaction giving an
anisotropic magnetic ordering. 

In this paper we focus on magnetism in the $\mathrm{U_{4}Ru_{7}Ge_{6}}$
compound. Although the first $\mathrm{U_{4}Ru_{7}Ge_{6}}$ single
crystals were grown already in late Eighties only a vague report on
ferromagnetism ($T_{\mathrm{C}}\sim\unit[7]{K}$, $\unit[0.2]{\mu_{\mathrm{B}}/U}$
ion in $\unit[5]{T}$ at $\unit[4.3]{K}$)\cite{RN4} and no information
on anisotropy can be found in literature. Therefore we have grown
a high quality single crystal of this compound and measured on it
the magnetization, AC susceptibility, thermal expansion, specific
heat and electrical resistivity with respect to temperature and magnetic
field. A weakly anisotropic ferromagnetism has been observed below
$T_{\mathrm{C}}=\unit[10.7]{K}$. The ferromagnetic ground state is
characterized by the easy-magnetization axis along the {[}111{]} crystallographic
direction. A magnetic phase transition at which the easy magnetization
axis changes from {[}111{]} to {[}001{]} is observed at $T_{r}=\unit[5.9]{K}$.
The {[}111{]} ({[}001{]}) phase is associated with a tiny rhombohedral
(tetragonal) distortion of the paramagnetic cubic crystal structure
which hardly can be observed by the X-ray diffraction but we demonstrate
that it can be indicated by thermal-expansion data. To shed more light
on microscopic mechanisms responsible for the $\mathrm{U_{4}Ru_{7}Ge_{6}}$
magnetism in some aspects very unusual for U compounds we also performed
first-principles electronic structure calculations.

\section{Experimental Details}

The $\mathrm{U_{4}Ru_{7}Ge_{6}}$ single crystal has been grown by
Czochralski method in a tri-arc furnace from constituent elements
(purity of Ru 3N5 and Ge 6N). The uranium metal was purified using
Solid State Electrotransport technique (SSE). Half of the crystal
was wrapped in the Ta foil (purity 4N), sealed in a quartz tube under
the vacuum of $\unit[1\cdot10^{-6}]{mbar}$ and subsequently annealed
at $\unit[1000]{\text{\textdegree C}}$ for 7 days. The high quality
of both crystals was verified by Laue diffraction using a Photonic
Science X-Ray Laue system with a CCD camera and single crystal X-ray
diffraction on a Rigaku R-Axis Rapid II diffractometer with a Mo lamp.
A part of each, the as-cast and annealed single crystal, respectively,
was pulverized and characterized by X-ray powder diffraction (XRPD)
at room temperature on a Bruker D8 Advance diffractometer with a Cu
lamp. The obtained data were evaluated by Rietveld technique\cite{RN55}
using FULLPROF/WINPLOTR software\cite{RN56,RN57}. The chemical composition
of each single crystal was verified by a scanning electron microscope
(SEM) Tescan Mira I LMH equipped with an energy dispersive X-ray (EDX)
detector Bruker AXS. Samples for individual experiments were afterward
cut from the annealed crystal with a fine wire saw to prevent induction
of additional stresses and lattice defects.

The magnetization (in a magnetic field up to $\unit[7]{T}$ applied
along the {[}001{]} and {[}111{]} directions, respectively) and AC
susceptibility (the AC magnetic field with the amplitude of $\unit[300]{\mu T}$
applied along {[}001{]}) were measured by a SQUID magnetometer in
MPMS 7T in the temperature range from $\unit[1.9]{K}$ to $\unit[300]{K}$.
The magnetization along {[}111{]} was measured up to $\unit[14]{T}$
by a VSM device at a PPMS 14T. Specific heat data were collected by
the thermal relaxation technique in the temperature range from $0.4$
to $\unit[20]{K}$ in the magnetic fields up to $\unit[9]{T}$ by
a PPMS 9T. The electrical resistivity measurements were realized with
the ACT option of the PPMS instruments with AC current applied along
{[}100{]} and {[}111{]}, respectively, in the temperature range from
$\unit[1.9]{K}$ to $\unit[300]{K}$. The MPMS 7T and both the PPMS
apparatuses are of Quantum Design Inc. production. The thermal expansion
measurements along the {[}100{]}, {[}001{]} and {[}111{]} directions,
respectively, in the temperature range from $\unit[1.9]{K}$ to $\unit[30]{K}$
using a miniature capacitance dilatometer\cite{RN50} implemented
in the PPMS 9T. All the instrumentation mentioned above is a part
of the Magnetism and Low Temperature Laboratories \textendash{} MLTL
(\url{http://mltl.eu/}). 

\section{Results}

\subsection{Crystal structure and chemical composition}

\begin{figure}
\begin{centering}
\includegraphics[width=8.6cm]{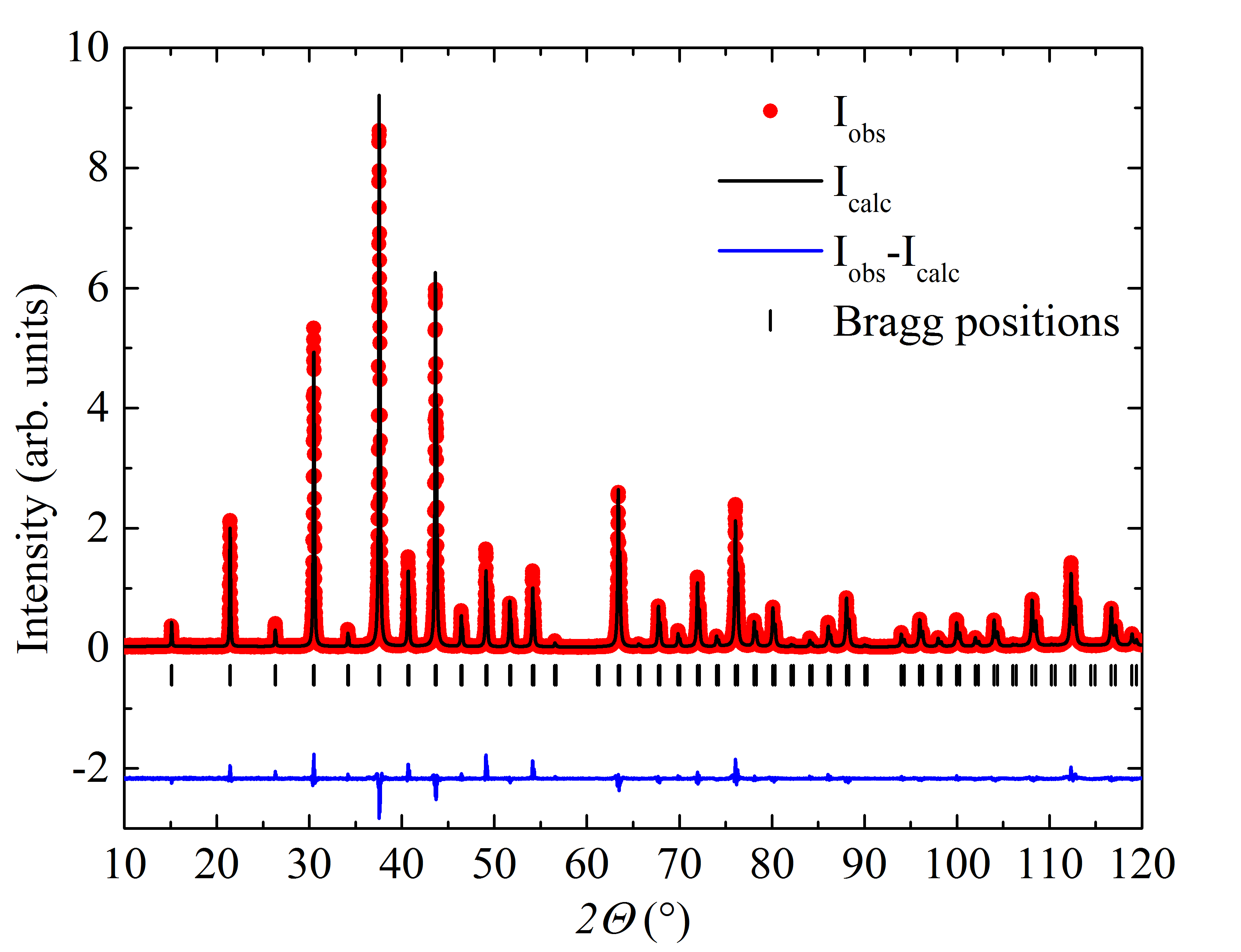}
\par\end{centering}
\caption{\label{fig:X-ray-powder-diffraction}X-ray powder diffraction pattern
of a pulverized $\mathrm{U_{4}Ru_{7}Ge_{6}}$ as cast single crystal. }

\end{figure}

The Rietveld refinement of XRPD data collected on the pulverized as-cast
(see the powder pattern in Figure \ref{fig:X-ray-powder-diffraction})
and annealed single crystal, respectively, confirmed previous reports\cite{RN4,RN5,RN8}
that $\mathrm{U_{4}Ru_{7}Ge_{6}}$ possesses at room temperature a
cubic crystal structure of the $Im\bar{3}m$ space group. The corresponding
lattice parameters (see Table \ref{fig:X-ray-powder-diffraction})
determined for the as cast and annealed crystal, respectively, are
nearly identical. 

\begin{table}
\begin{centering}
\begin{tabular*}{8.6cm}{@{\extracolsep{\fill}}ccc}
\toprule 
Space group $Im\bar{3}m$  & as-cast & annealed\tabularnewline
\midrule
\midrule 
$a$ & $\unit[8.2934(2)]{\textrm{Å}}$ & $\unit[8.2933(3)]{\textrm{Å}}$\tabularnewline
\midrule
$\mathrm{U}(x,y,z),\thinspace8c$ & (0.25, 0.25, 0.25) & (0.25, 0.25, 0.25)\tabularnewline
$\mathrm{Ru_{1}}(x,y,z),\thinspace12d$ & (0.25, 0, 0.5) & (0.25, 0, 0.5)\tabularnewline
$\mathrm{Ru_{2}}(x,y,z),\thinspace2a$ & (0, 0, 0) & (0, 0, 0)\tabularnewline
$\mathrm{Ge}(x,y,z),\thinspace12e$ & (0.31375(13), 0, 0) & (0.31360(11), 0, 0)\tabularnewline
\end{tabular*}
\par\end{centering}
\caption{\label{tab:Results-of-structure}Results of structure analysis.}
\end{table}

Elemental mapping by EDX confirmed homogeneity of all the studied
samples. The average of multiple point scans from the different part
of the sample gives resulting stoichiometry 4.4(4):6.9(2):5.7(1).
It points to a slight Ge deficiency.

\subsection{Low temperature magnetization}

\begin{figure}
\begin{centering}
\includegraphics[width=8.6cm]{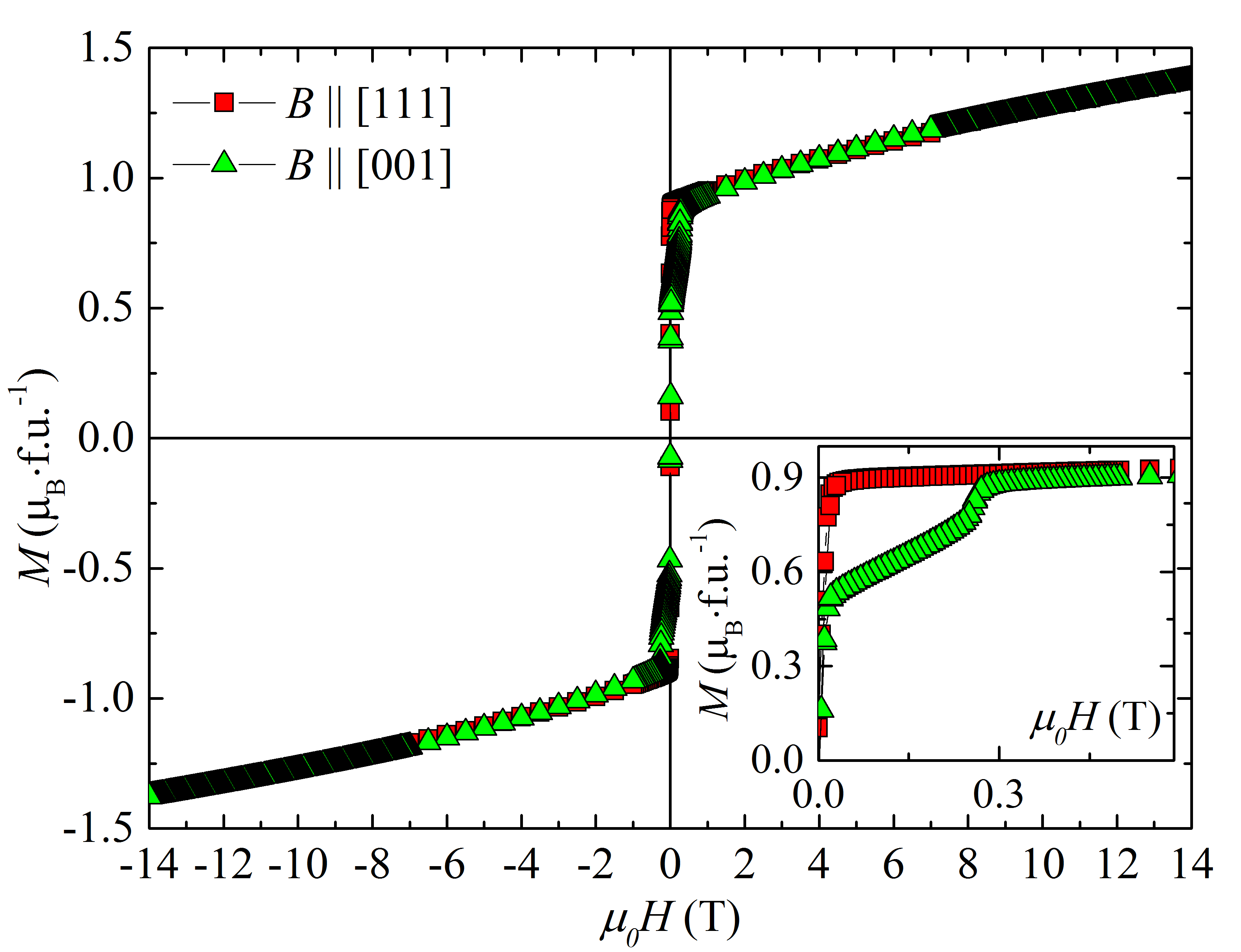}
\par\end{centering}
\caption{\label{fig:Field-dependence-of}Field dependence of magnetization
isotherms of $\mathrm{U_{4}Ru_{7}Ge_{6}}$ at $\unit[1.9]{K}$ for
{[}111{]}, and {[}001{]} directions of applied magnetic field. Inset
shows a low field detail.}
\end{figure}

The magnetization curves $M(\mu_{0}H)$ measured at $\unit[1.9]{K}$
in a magnetic field applied along the {[}001{]} and {[}111{]} direction,
respectively, document that $\mathrm{U_{4}Ru_{7}Ge_{6}}$ is at this
temperature ferromagnetic with the easy magnetization direction along
the {[}111{]} axis (see Figure \ref{fig:Field-dependence-of}). The
spontaneous magnetization $M_{\mathrm{S\left[111\right]}}=\unit[0.88\left(1\right)]{\mu_{B}}$
(obtained as an extrapolation of the $M_{\left[111\right]}\left(\mu_{0}H\right)$
dependence to $\mu_{0}H=\unit[0]{T}$). The spontaneous magnetization
value along the {[}001{]} direction is lower, $M_{\mathrm{S\left[001\right]}}=\unit[0.51\left(1\right)]{\mu_{B}}$,
respectively. This value is in very good agreement with Néel phase
law\cite{RN51} for a cubic system. It proposes that $M_{\mathrm{S\left[001\right]}}$
can be obtained from the $M_{\mathrm{S\left[111\right]}}$ value multiplied
by the appropriate direction cosine \textendash{} i.e. $M_{\mathrm{S\left[001\right]}}=\sqrt{\frac{1}{3}}M_{\mathrm{S\left[111\right]}}$.
For fields higher than $\unit[0.3]{T}$ the $M_{\left[111\right]}(\mu_{0}H)$
and $M_{\left[001\right]}(\mu_{0}H)$ curves merge (we estimate the
$\mu_{0}H_{\mathrm{a}}$ value $\unit[300]{mT}$), however, do not
saturate in the field up to the $\unit[14]{T}$ where the magnetization
reaches the value of $\unit[1.4]{\mu_{B}/f.u.}$. The poor saturation
of the magnetization in high magnetic fields is typical for itinerant
electron ferromagnets. The increasing magnetic moment with increasing
the magnetic field is then reflecting the magnetic-field induced change
of electronic structure (additional splitting of the majority and
minority subbands). 

The observed very weak magnetization anisotropy makes $\mathrm{U_{4}Ru_{7}Ge_{6}}$
a unique exception among the U ferromagnets which are by rule strongly
anisotropic with anisotropy fields of several hundred T\cite{RN20}.
In the next chapter we approach this issue by the first-principles
electronic structure calculations focused on the microscopic origin
of U magnetic moments and magnetocrystalline anisotropy. 

\subsection{First-principles calculations}

\begin{figure}
\begin{centering}
\includegraphics[width=8.6cm]{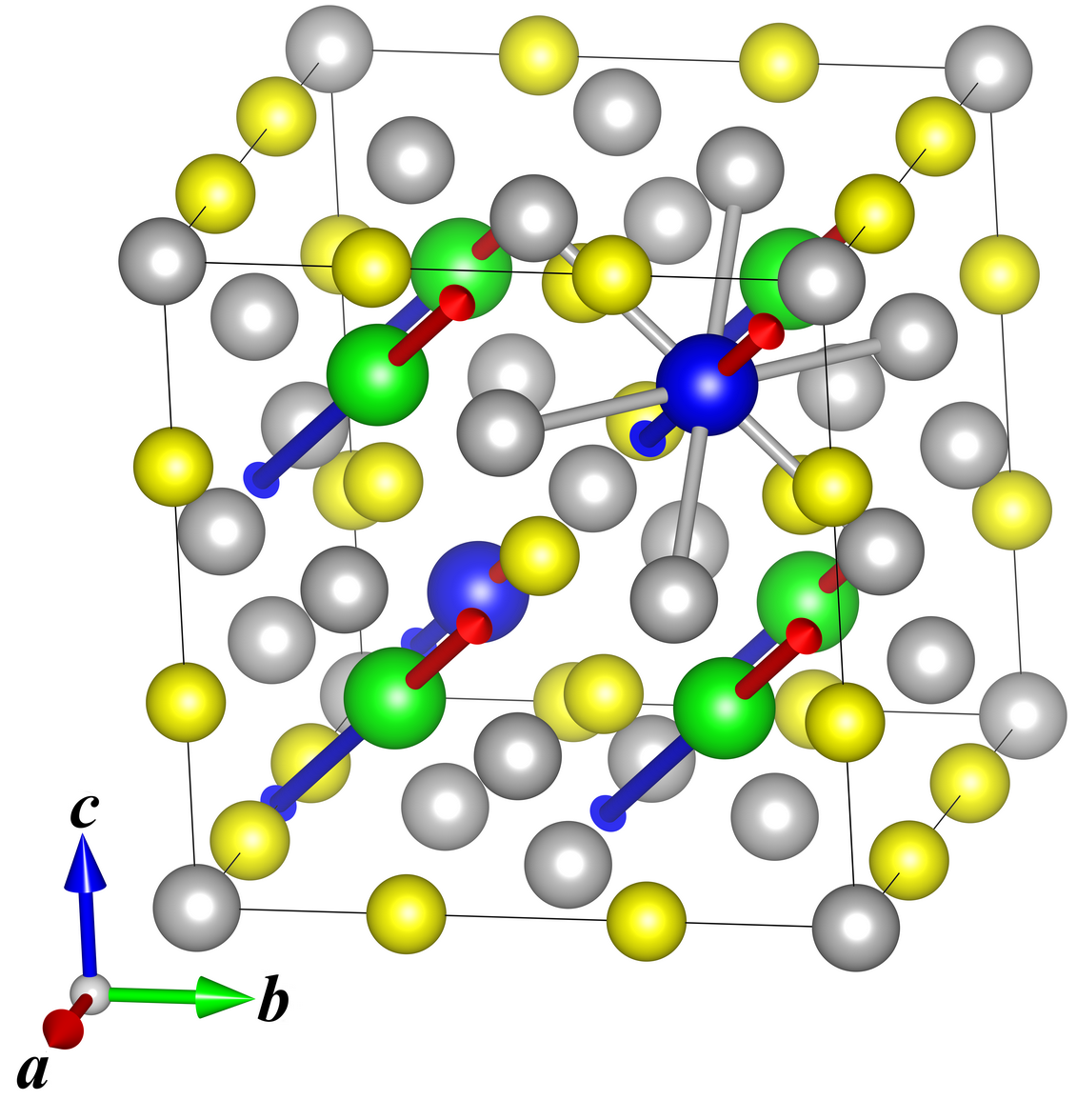}
\par\end{centering}
\caption{\label{fig:Structure-of-}Structure of $\mathrm{U_{4}Ru_{7}Ge_{6}}$
together with magnetic moments from theoretical calculations. Arrows
for magnetic moments are in proper relative scale but in arbitrary
units. $\mathrm{U_{1}}$ ions are blue, $\mathrm{U_{2}}$ are green,
$\mathrm{Ru_{3}}$ and $\mathrm{Ru_{4}}$ are gray and $\mathrm{Ge_{5}}$
are yellow.}
\end{figure}

To obtain microscopic information about magnetism in $\mathrm{U_{4}Ru_{7}Ge_{6}}$
we applied the first-principles methods based on the density functional
theory (DFT). The Kohn-Sham-Dirac 4-components equations have been
solved by using the latest version of the full-potential-local-orbitals
(FPLO) computer code\cite{RN58}. Several $k$-meshes in the Brillouin
zone were involved to ensure the convergence of charge densities,
total energy and magnetic moments. For the sake of simplicity we assumed
a collinear ferromagnetic structure. In $\mathrm{U_{4}Ru_{7}Ge_{6}}$
the total ground-state magnetic moment has been found pointing along
the {[}111{]} direction. Due to the s-o interaction the symmetry is
reduced from 48 to 12 symmetry operations. Instead of four symmetrically
equivalent U ion sites in the scalar relativistic treatment with a
spin-only magnetic moment we have the spin and orbital angular momenta
coupled by the relativistic s-o interaction which divides the U ions
in two subgroups consistently with the expected rhombohedral distortion
induced by magnetoelastic interaction in case of the {[}111{]} easy
magnetization direction. The U ions in the first subgroup at the $\mathrm{U_{1}}$
positions (0.25, 0.25, 0.25) and (0.75, 0.75, 0.75), respectively
have a very small total magnetic moment of $\unit[0.01]{\mu_{\mathrm{B}}}$
due to cancellation of the almost equal-size antiparallel spin and
orbital moment (see Figure \ref{fig:Structure-of-}). The second subgroup
includes the U ions at the remaining six $\mathrm{U_{2}}$ positions
bearing the spin magnetic moment of $\unit[-0.56]{\mu_{\mathrm{B}}}$
and the orbital magnetic moment of $\unit[0.79]{\mu_{\mathrm{B}}}$.
There are also some hybridization-induced Ru magnetic moments, which
summed up give $\unit[0.29]{\mu_{\mathrm{B}}}$, whereas the calculated
Ge magnetic moment is negligible. The summation over the seventeen
ions (one formula unit) in the primitive crystallographic cell gives
the total magnetic moment of $\unit[1.01]{\mu_{\mathrm{B}}}$ which
is somewhat larger than the spontaneous magnetic moment determined
by experiment.

We have also performed relativistic calculation with the moment along
{[}001{]}. In this case the cubic symmetry is reduced to tetragonal
and all U moments have the same value of spin ($\unit[-0.529]{\mu_{\mathrm{B}}}$)
and orbital ($\unit[0.706]{\mu_{\mathrm{B}}}$) component. 

When calculating the magnetocrystalline anisotropy energy between
the configurations of magnetic moments aligned along the {[}111{]}
and {[}001{]} axis the values of the total energy were used. In agreement
with experiment we have found the ground state with the total moment
pointing to the {[}111{]} direction. The excited state with the moment
pointing to the {[}001{]} direction is $\unit[0.9]{meV}$ above the
ground state. This value is as to order of magnitude in agreement
with the energy corresponding to $T_{\mathrm{r}}$, the temperature
of {[}111{]} to {[}100{]} spin reorientation transition.

\subsection{Magnetization near the phase transitions}

\begin{figure}
\begin{centering}
\includegraphics[width=8.6cm]{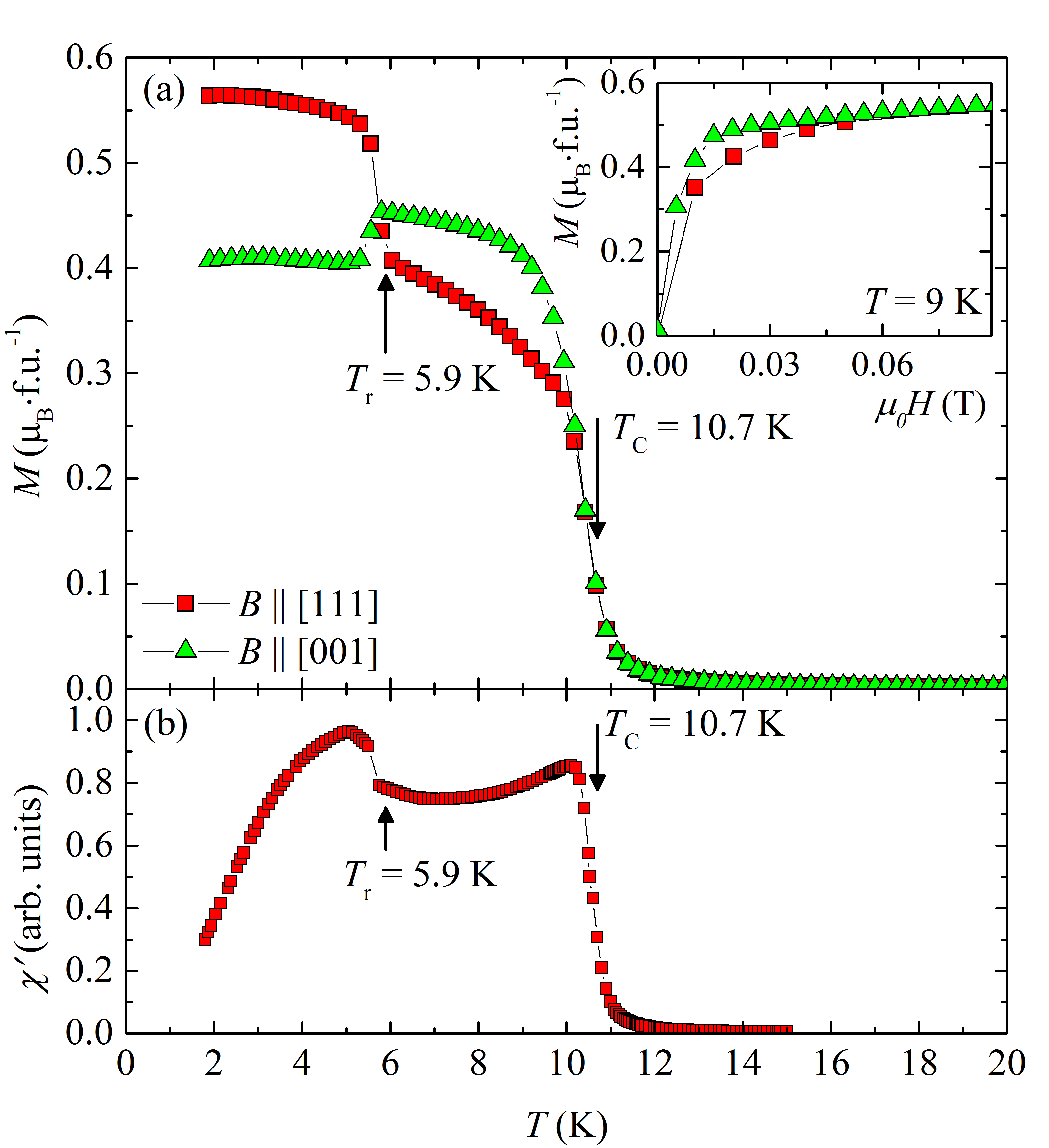}
\par\end{centering}
\caption{\label{fig:Temperature-dependence-of}Temperature dependence of the
magnetization of $\mathrm{U_{4}Ru_{7}Ge_{6}}$ measured in the magnetic
field of $\unit[10]{mT}$ (in the ZFC regime) applied along the {[}111{]}
and {[}100{]} direction, respectively (a) and temperature dependence
of the AC susceptibility in the AC field applied along the {[}111{]}
(b). Inset in the upper panel shows field dependence of magnetization
at $\unit[9]{K}$.}

\end{figure}

The Curie temperature, $T_{\mathrm{C}}$, of a ferromagnet is frequently
estimated as the temperature of the inflection point of the $M$ vs.
$T$ curve measured in a low magnetic field and/or of the temperature
dependence of the AC-susceptibility. In Figure \ref{fig:Temperature-dependence-of}
(a) one can see that the temperature dependences of magnetization
$M(T)$ measured in the external magnetic field of $\unit[10]{mT}$
applied along the {[}111{]} and {[}100{]} direction, respectively,
have one inflection point at the same temperature of $\unit[10.6\pm0.1]{K}$
($\sim T_{\mathrm{C}}$) and are crossing at $T_{r}=\unit[5.9\pm0.1]{K}$.
Two sharp anomalies at corresponding temperatures can be seen in the
$\chi_{\mathrm{AC}}(T)$ dependence seen in the Figure \ref{fig:Temperature-dependence-of}
(b). These results clearly document that $\mathrm{U_{4}Ru_{7}Ge_{6}}$
orders at $T_{\mathrm{C}}$ ferromagnetically with the easy magnetization
axis {[}100{]} (which is demonstrated by the $\unit[9]{K}$ magnetization
isotherms in the inset of Figure \ref{fig:Temperature-dependence-of}
(a)). At $T_{\mathrm{r}}$ it undergoes a spin reorientation transition
to the ground state characterized by the easy magnetization axis along
{[}111{]}.

\begin{figure}
\begin{centering}
\includegraphics[width=8.6cm]{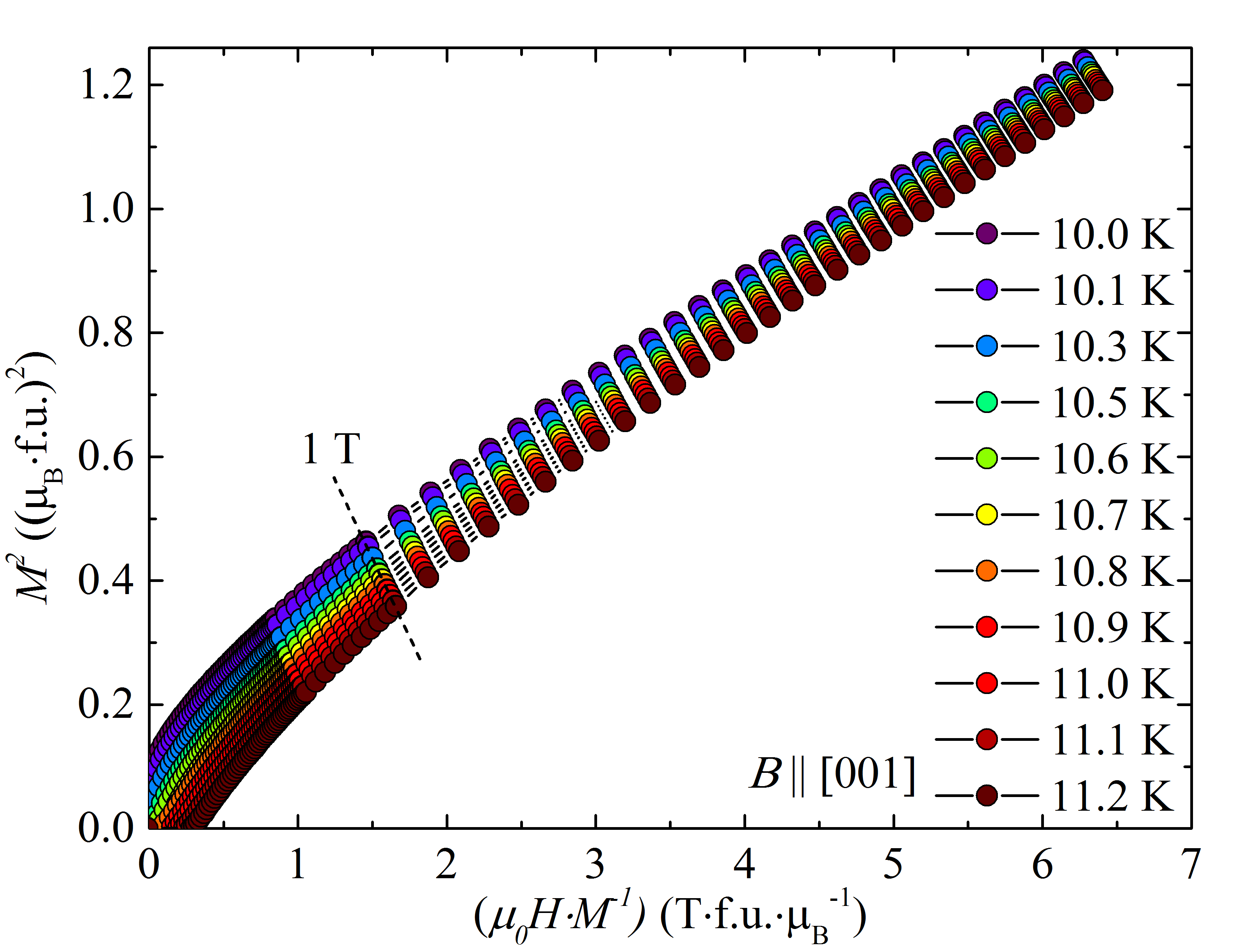}
\par\end{centering}
\caption{\label{fig:Arrott-plots-for}Arrott plots for $\mathrm{U_{4}Ru_{7}Ge_{6}}$
in the magnetic field applied along the {[}001{]} direction.}

\end{figure}

A rigorous way of determining Curie temperature of a ferromagnet from
magnetization data is based on the analysis of Arrott plots ($M^{2}$
vs. $\mu_{0}H/M$)\cite{RN13}. Linear Arrott plots are in fact a
graphical representation of the Ginsburg-Landau mean field theory
of magnetism in the vicinity of the ferromagnetic to paramagnetic
second order phase transition. In Figure \ref{fig:Arrott-plots-for}
one can see that the Arrott plots for $\mathrm{U_{4}Ru_{7}Ge_{6}}$
in the magnetic field applied along the easy magnetization direction
{[}001{]} are almost linear with varying slope for magnetic fields
between 1 and $\unit[7]{T}$ whereas for lower fields they became
slightly concave. The linear extrapolations of high-field data to
the vertical axis are marking the values of $M^{2}$, which are considered
as estimates of the square spontaneous magnetization $M_{\mathrm{S}}^{2}$.
The spontaneous magnetization as the order parameter of a ferromagnetic
phase vanishes at $T_{\mathrm{C}}$. Note that the use of the linear
extrapolations from high fields in case of the concave curvature of
$\mathrm{U_{4}Ru_{7}Ge_{6}}$ Arrott plots in low fields leads to
a certain overestimation of $M_{\mathrm{S}}$ values and consequently
to a higher estimated $T_{\mathrm{C}}$ value.

\begin{figure}
\begin{centering}
\includegraphics[width=8.6cm]{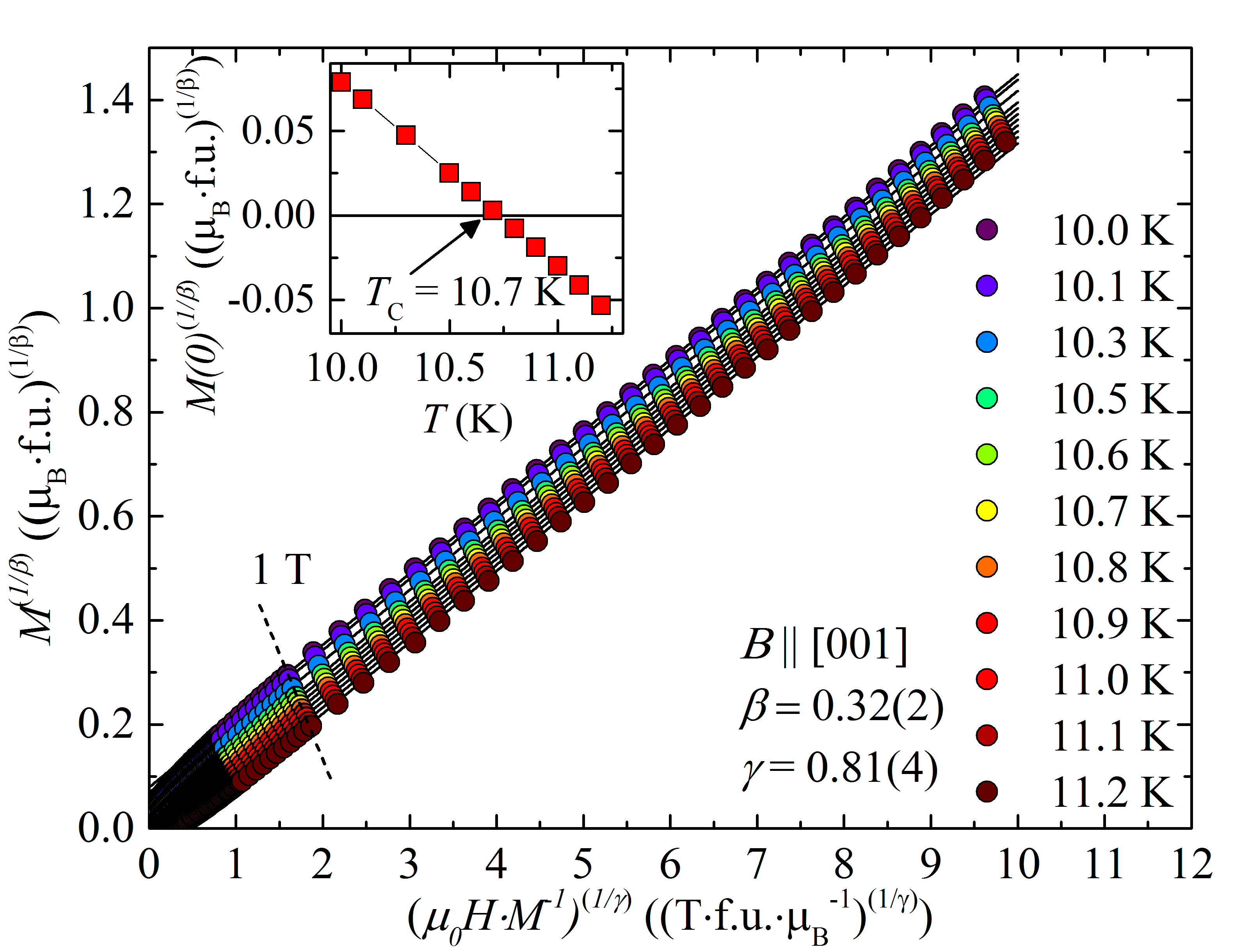}
\par\end{centering}
\caption{\label{fig:Arrott-Noakes-plots-reflecting}Arrott-Noakes plots reflecting
the equation of states (with $\beta=0.31\pm0.03$ and $\gamma=0.81\pm0.04$)
for $\mathrm{U_{4}Ru_{7}Ge_{6}}$ in the magnetic field applied along
the {[}001{]} direction. }

\end{figure}

A more precise $T_{\mathrm{C}}$ value may be expected from the generalized
approach using the Arrott-Noakes equation of state $\left(\mu_{0}H/M\right)^{1/\gamma}=\left(T-T_{\mathrm{C}}\right)/T_{1}+\left(M/M_{1}\right)^{1/\beta}$,
where $M_{1}$ and $T_{1}$ are material constants\cite{RN12}. We
re-analyzed our data by plotting them as $M^{1/\beta}$ vs. $\left(\mu_{0}H/M\right)^{1/\gamma}$
while $\beta$ and $\gamma$ values were chosen to get the best possible
linearity of these plots keeping them parallel with constant slope.
The values $\beta=0.31\pm0.03$ and $\gamma=0.81\pm0.04$ lead to
a linear dependence in the broad field range except the very low fields.
This construction is plotted in Figure \ref{fig:Arrott-Noakes-plots-reflecting}
for all measured isotherms. This approach leads to $T_{\mathrm{C}}=\unit[10.7\pm0.1]{K}$
(see inset of Figure \ref{fig:Arrott-Noakes-plots-reflecting}) that
is in agreement with the estimated value from the low field magnetization
data.

The critical exponent $\delta$ in ideal case might satisfy the Widom
scaling relation $\delta=1+\gamma/\beta$ (Ref. \cite{RN14}) which
for $\beta$ and $\gamma$, provided by the Arrott-Noakes analysis,
gives $\delta=3.52\pm0.04$ that is in excellent agreement with the
value $\delta=3.55\pm0.04$ obtained from direct fitting of the critical
isotherm that should follow $M\sim\left(\mu_{0}H\right)^{1/\delta}$.
Note that the critical exponents for the mean field approximation
are $\beta=0.5$, $\gamma=1$ and $\delta=3$ (Ref. \cite{RN53}),
however, we should mention the work of Yamada\cite{RN52} who has
shown that the spin fluctuations in weak itinerant ferromagnets lead
to the Arrott plots linear in strong magnetic fields and bent downwards
at the region of small magnetizations.

\subsection{Paramagnetic susceptibility}

\begin{figure}
\begin{centering}
\includegraphics[width=8.6cm]{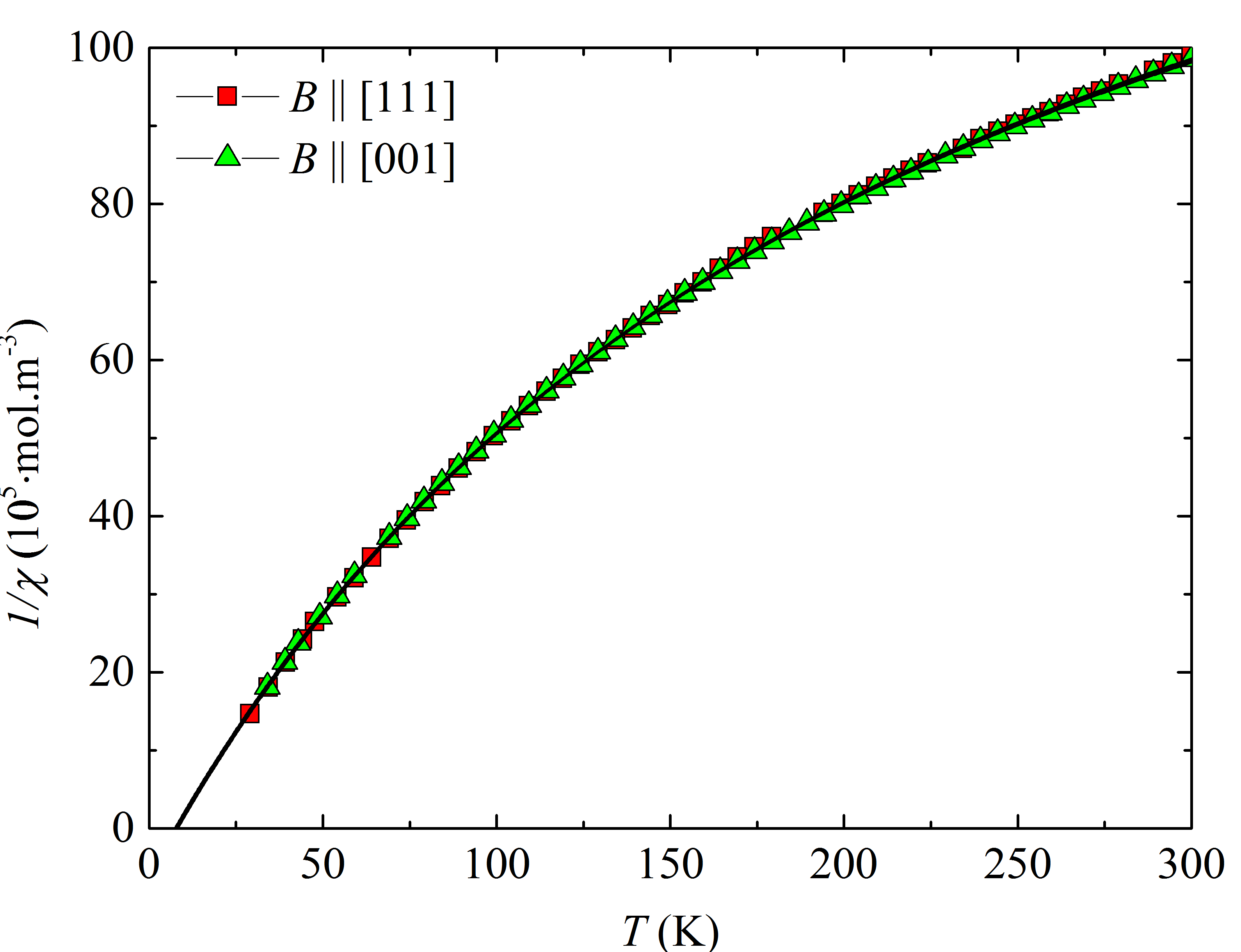}
\par\end{centering}
\caption{\label{fig:Temperature-dependence-of-1}Temperature dependence of
inverse susceptibility of $\mathrm{U_{4}Ru_{7}Ge_{6}}$ in the magnetic
field of $\unit[1]{T}$ applied along the {[}111{]} and {[}001{]}
direction. The full curve represents the fit with a modified Curie-Weiss
law. }

\end{figure}

The nearly identical temperature dependences of paramagnetic susceptibility
measured along the {[}111{]} and {[}100{]} direction, respectively
document isotropic paramagnetic state of $\mathrm{U_{4}Ru_{7}Ge_{6}}$
(see the $1/\chi$ vs. $T$ plot in Figure \ref{fig:Temperature-dependence-of-1}).
The susceptibility values at temperatures above $\unit[30]{K}$ can
be well fitted with a modified Curie-Weiss law in temperature range
with parameters shown in Table \ref{tab:Results-of-the}.

\begin{table}
\begin{centering}
\begin{tabular*}{8.6cm}{@{\extracolsep{\fill}}cccc}
\toprule 
$B\parallel$ & $\unit[\mu_{\mathrm{eff}}]{\left(\mu_{\mathrm{B}}/\mathrm{U}\right)}$ & $\unit[\theta_{\mathrm{P}}]{(\mathrm{K})}$ & $\unit[\chi_{0}\cdot10^{-8}]{\left(m^{3}\cdot mol^{-1}\right)}$\tabularnewline
\midrule
\midrule 
{[}001{]} & 1.43 & 7.5 & 5.8\tabularnewline
\midrule 
{[}111{]} & 1.43 & 8.1 & 5.7\tabularnewline
\bottomrule
\end{tabular*}
\par\end{centering}
\caption{\label{tab:Results-of-the}Results of the modified Curie-Weiss fit
in the temperature interval $\unit[30-300]{K}$.}

\end{table}

\subsection{Heat capacity}

\begin{figure}
\begin{centering}
\includegraphics[width=8.6cm]{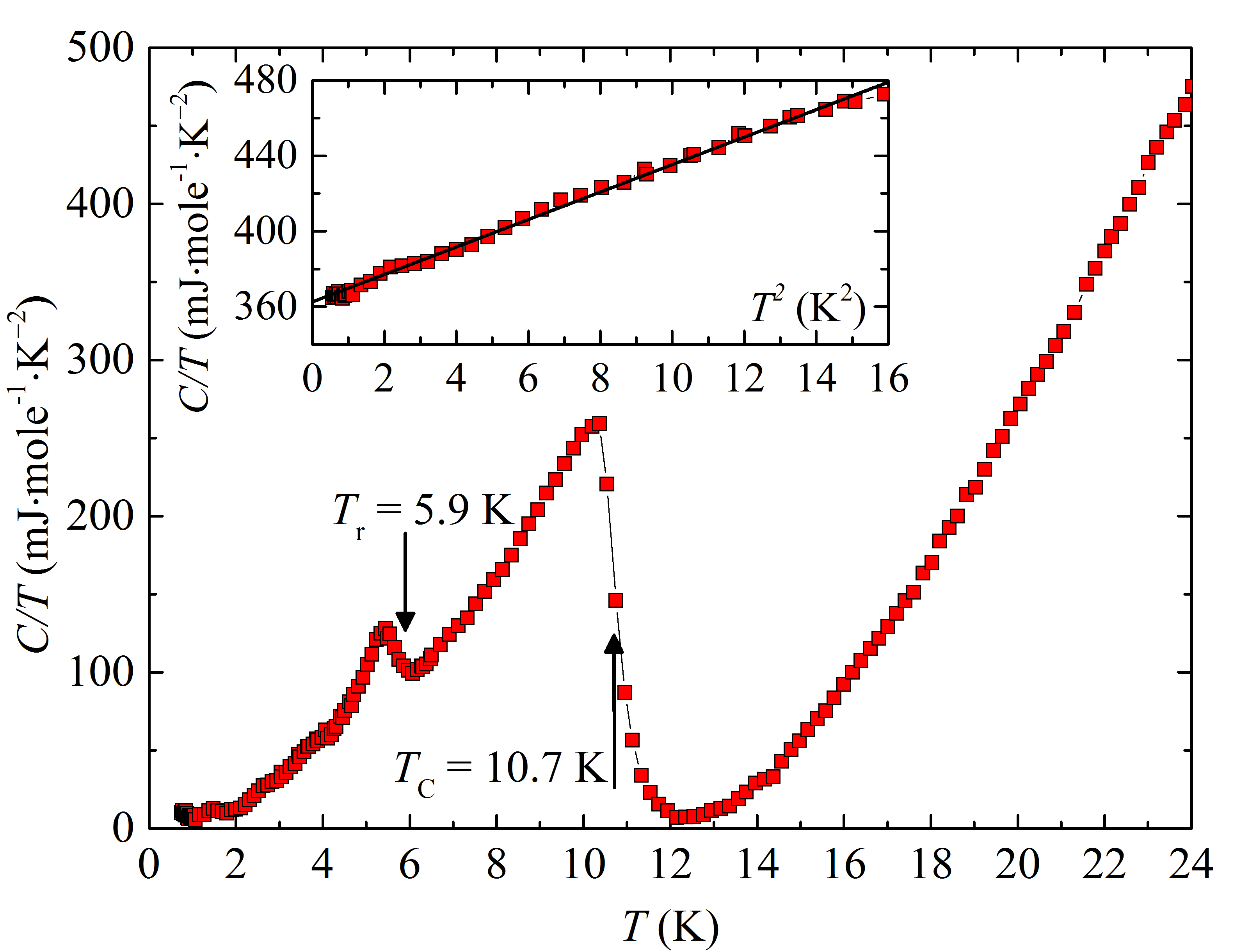}
\par\end{centering}
\caption{\label{fig:Temperature-dependence-of-2}Temperature dependence of
the heat capacity ($C/T$ vs. $T$ plot) of $\mathrm{U_{4}Ru_{7}Ge_{6}}$.
Inset: the $C/T$ vs. $T^{2}$ plot.}

\end{figure}

Heat capacity data show a clear anomaly at $\unit[10.7]{K}$ as displayed
in Figure \ref{fig:Temperature-dependence-of-2}. This temperature
coincides with the $T_{\mathrm{C}}$ value determined from magnetization
data by Arrott-Noakes plot analysis. The estimated magnetic entropy
at $T_{\mathrm{C}}$ (i.e. integrated from $\unit[0.3]{K}$ to $T_{\mathrm{C}}$)
of $0.2\cdot\mathrm{R}\ln2$ is much lower than $\mathrm{R\ln2}$
which is another evidence of itinerant electron magnetism in $\mathrm{U_{4}Ru_{7}Ge_{6}}$.
The magnetic moment reorientation transition at $T_{\mathrm{r}}$
is reflected in a tiny but clear peak at this temperature.

The gamma coefficient of the electronic specific heat determined from
a standard $C/T$ vs. $T^{2}$ plot (see Inset in Figure \ref{fig:Temperature-dependence-of-2})
constructed from data below $\unit[4]{K}$ amounts to $\unit[362]{mJ\cdot mol^{-1}\cdot K^{-2}}$.
The value related to one U ion $=\unit[90.5]{mJ\cdot mol^{-1}\cdot K^{-2}}$
reflects presence of the U 5$f$-electron states at $E_{\mathrm{F}}$
similar to numerous other U intermetallics with itinerant 5$f$-electrons
which usually exhibit elevated values somewhere between 30 and $\unit[100]{mJ\cdot mol^{-1}\cdot K^{-2}}$
per U ion. 

\subsection{Thermal expansion and magnetostriction}

\begin{figure}
\begin{centering}
\includegraphics[width=8.6cm]{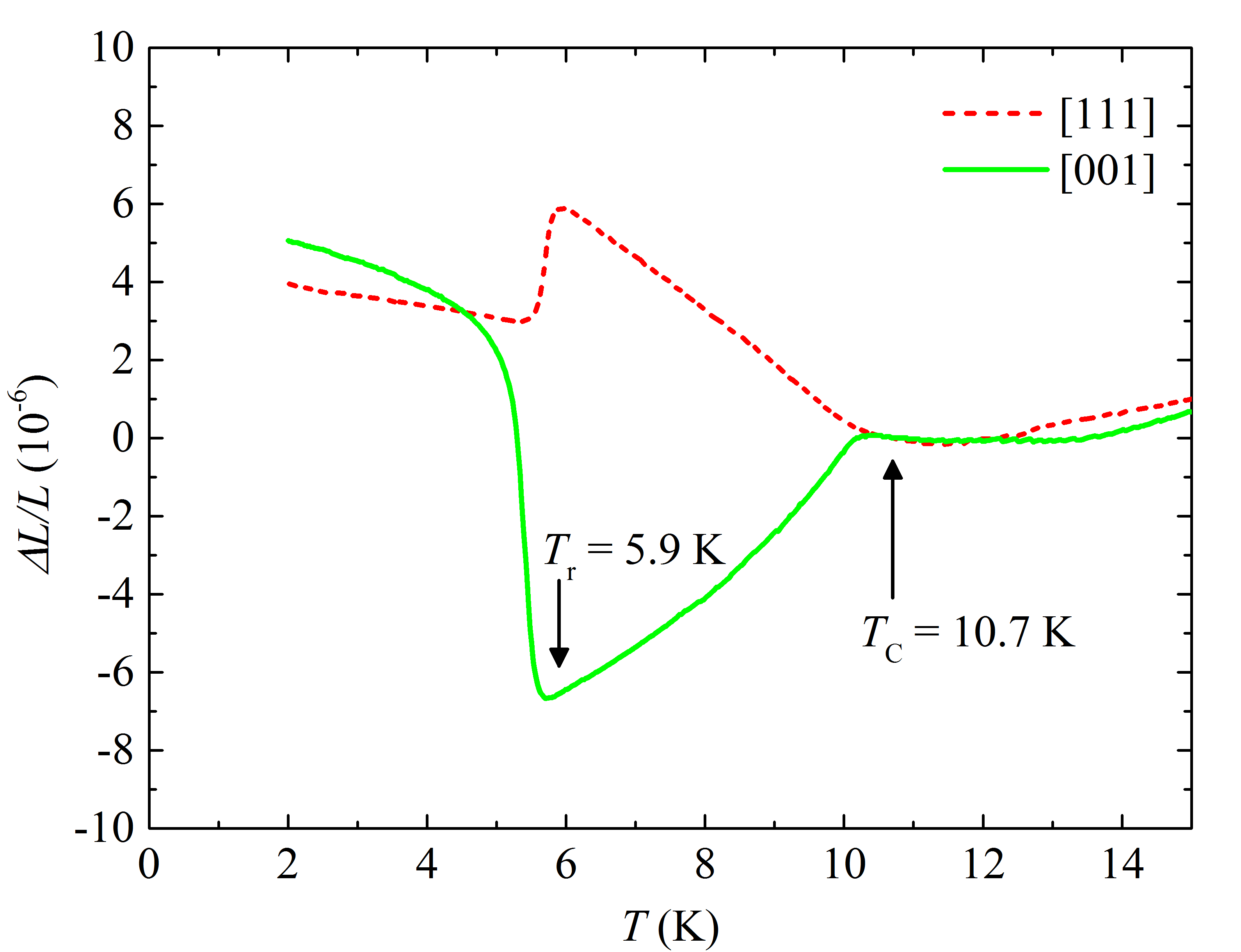}
\par\end{centering}
\caption{\label{fig:Linear-thermal-expansion}Linear thermal expansion of $\mathrm{U_{4}Ru_{7}Ge_{6}}$
along the {[}001{]} and {[}111{]} directions, respectively.}

\end{figure}

The linear thermal expansion $\Delta L/L$ measured along the {[}001{]}
({[}111{]}) direction in zero magnetic field which can be seen in
Figure \ref{fig:Linear-thermal-expansion}, shows two distinct anomalies
which can be attributed to the magnetic phase transitions revealed
by magnetization measurements. When cooling from higher temperatures
a downturn (upturn) having onset in the vicinity of $T_{\mathrm{C}}$
is followed by a steep increase (decrease) of the corresponding $\Delta L/L$
below $\sim\unit[6]{K}$ ($\sim T_{\mathrm{r}}$). 

Most of the thermal expansion studies of ferromagnets revealing the
spontaneous magnetostriction at temperatures $T<T_{\mathrm{C}}$ were
done by using the X-ray or neutron diffraction\cite{RN59} . We investigated
the crystal structure of $\mathrm{U_{4}Ru_{7}Ge_{6}}$ by X-ray powder
diffraction at low temperatures down to $\unit[3]{K}$. No change
of diffraction within the experimental error data has been observed
below $\unit[11]{K}$.

From Figure \ref{fig:Linear-thermal-expansion} it is however evident
that our thermal expansion data obtained by dilatometer on the $\mathrm{U_{4}Ru_{7}Ge_{6}}$
single crystal clearly demonstrate the existence of lattice distortions
in ferromagnetic state. The distortions here are, however, very small
($<10^{-5}$).

The dilatometer enables us to determine the crystal distortions along
the 3 perpendicular crystal axes {[}100{]}, {[}010{]} and {[}001{]},
respectively. To study the corresponding linear spontaneous magnetostriction
of a ferromagnet with a dilatometer one should perform measurements
on a single crystal sample containing only one ferromagnetic domain.
Knowing magnetization data we studied the phase with the easy magnetization
along {[}001{]} stable at temperatures $T_{\mathrm{r}}<T<T_{\mathrm{C}}$
in the following way. We first cooled the crystal down to $\unit[6.2]{K}$
(the temperature just above the onset of the spin reorientation transition)
in the field of $\unit[1]{T}$ parallel to the {[}001{]} direction,
at this temperature decreased the magnetic field down to $\unit[30]{mT}$
and then measured the thermal expansion in the longitudinal $\left(\Delta L/L\right)_{\left[001\right]}$
and transversal $\left(\Delta L/L\right)_{\left[100\right]}$ geometry,
respectively, with increasing temperature up to $T_{\mathrm{C}}$.
In this experiment the thermal expansion along the $c$- and $a$-axis,
respectively, of the ferromagnetic tetragonally distorted structure
was measured. The field of $\unit[30]{mT}$ applied along {[}001{]}
is the minimum field maintaining the single-domain sample with the
magnetic moment oriented along the $c$-axis. These $\left(\Delta L/L\right)_{\left[001\right]}$
and $\left(\Delta L/L\right)_{\left[100\right]}$ thermal expansion
data can be taken as a reasonable approximation of spontaneous linear
magnetostriction (denoted as $\lambda_{\mathrm{S\left[001\right]}}$
and $\lambda_{\mathrm{S\left[100\right]}}$, respectively).

\begin{figure}
\begin{centering}
\includegraphics[width=8.6cm]{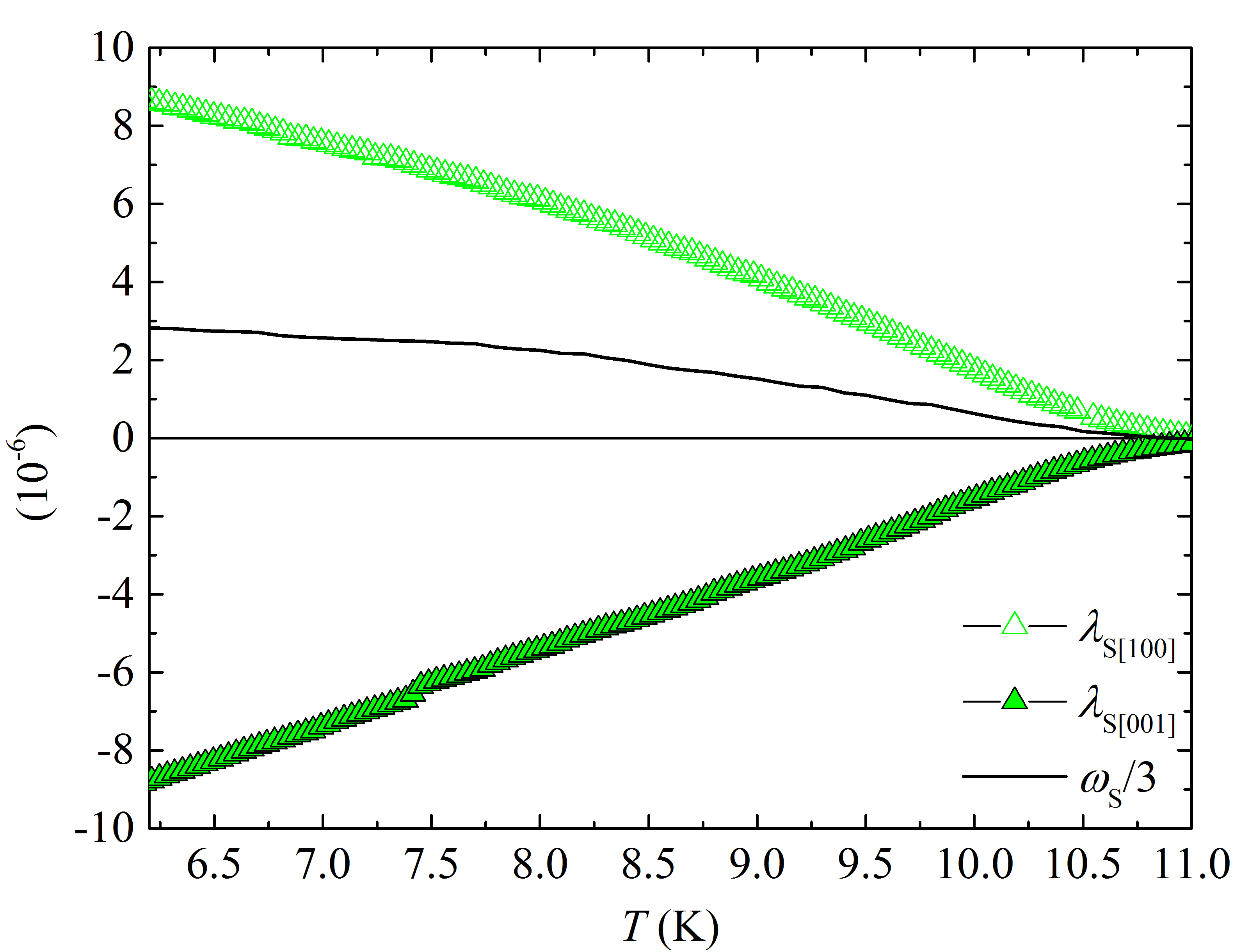}
\par\end{centering}
\caption{\label{fig:Thermal-linear-expansion}Linear thermal expansion of $\mathrm{U_{4}Ru_{7}Ge_{6}}$
along the {[}001{]} and {[}100{]} direction in the magnetic field
of $\unit[30]{mT}$ applied along {[}001{]} and the corresponding
thermal volume expansion at temperatures from $\unit[6.2]{K}$ to
$T_{\mathrm{C}}$. The presented data are considered as the best estimation
of corresponding spontaneous magnetostriction (see text). }

\end{figure}

One can see that the values of the spontaneous tetragonal distortion
is negative (the $c$-axis shrinks) and very small ($\lambda_{\mathrm{S\left[001\right]}}\sim-9\cdot10^{-6}$
at $\unit[6.2]{K}$). Simultaneously the $a$-axis expands ($\lambda_{\mathrm{S\left[100\right]}}=\lambda_{\mathrm{S\left[010\right]}}\sim9\cdot10^{-6}$
at $\unit[6.2]{K}$). Figure \ref{fig:Thermal-linear-expansion} shows
also the corresponding spontaneous thermal volume expansion usually
denoted as $\omega_{\mathrm{S}}\left(=2\cdot\lambda_{\mathrm{S\left[100\right]}}+\lambda_{\mathrm{S}\left[001\right]}\right)$.
As expected for an itinerant electron ferromagnet the volume of the
$\mathrm{U_{4}Ru_{7}Ge_{6}}$ lattice expands below $T_{\mathrm{C}}$.
Reasonable data can be collected only down to $\unit[6.2]{K}$. Below
this temperature the magnetic-moment-reorientation transition from
{[}001{]} to {[}111{]} direction commences, the direction of magnetic
moment becomes uncertain despite the magnetic field of $\unit[30]{mT}$
is applied along {[}001{]}.

\subsection{Electrical resistivity}

\begin{figure}
\begin{centering}
\includegraphics[width=8.6cm]{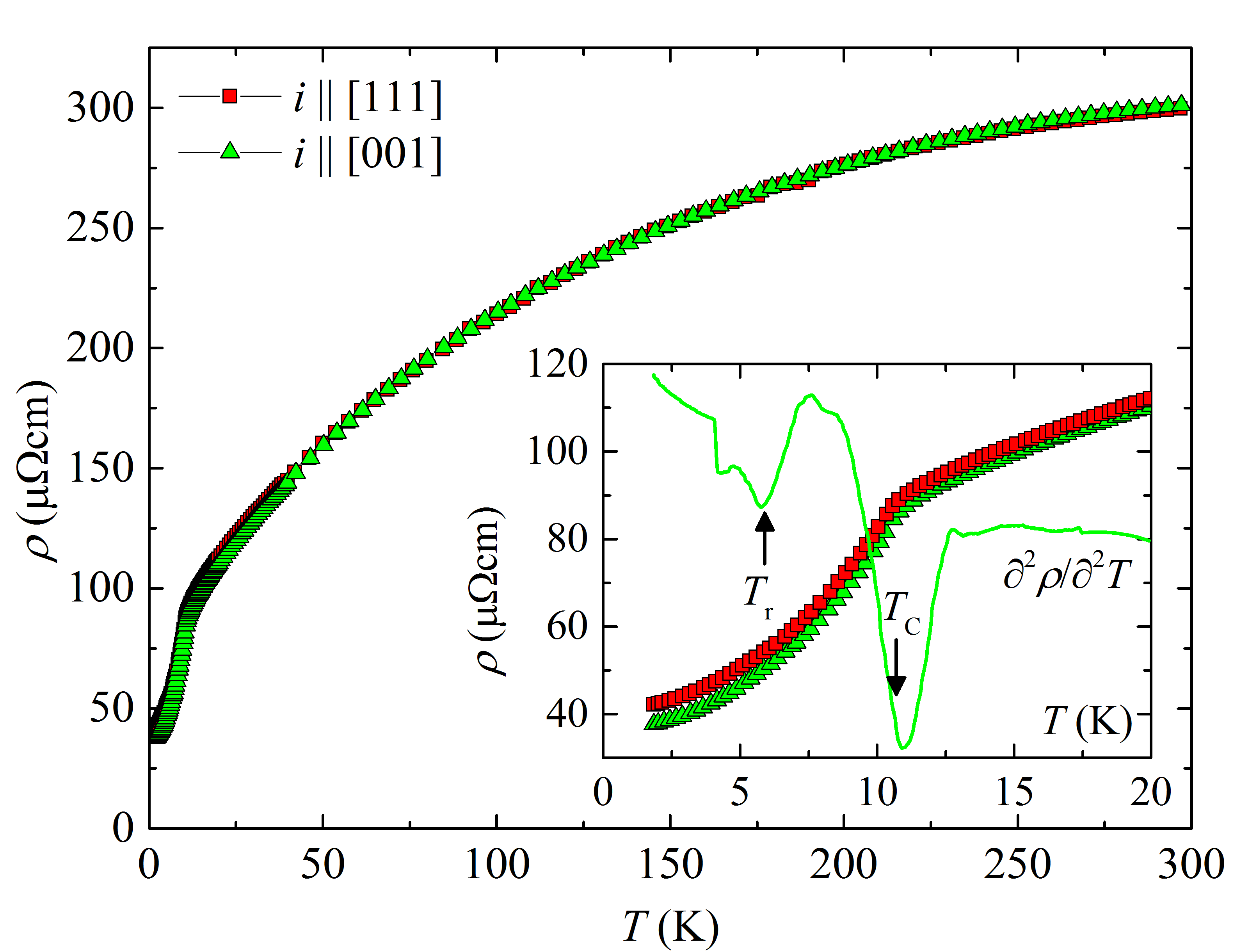}
\par\end{centering}
\caption{\label{fig:Temperature-dependence-of-3}Temperature dependence of
electrical resistivity of the $\mathrm{U_{4}Ru_{7}Ge_{6}}$ measured
for current along {[}111{]} and {[}100{]} direction.}

\end{figure}

The almost identical corresponding values of the electrical resistivity
measured for current along {[}111{]} and {[}100{]} direction, respectively,
over the entire temperature range $\unit[2-300]{K}$ indicates a quite
isotropic electron transport in $\mathrm{U_{4}Ru_{7}Ge_{6}}$ (see
Figure \ref{fig:Temperature-dependence-of-3}). The downward concave
$\rho\left(T\right)$ curve in the paramagnetic state resembles rather
the behavior of transition metal compounds. This trend changes at
$T_{\mathrm{C}}$ to the low-temperature upward concave curve with
the $T^{2}$ scaling. When inspecting the second derivative we observe
a deep minimum of $\partial^{2}\rho/\partial^{2}T$ at $T_{\mathrm{C}}$
and a local minimum at $T_{\mathrm{r}}$. 

\section{Discussion}

The $M(\mu_{0}H)$ curve measured for the hard magnetization direction
{[}100{]} merges with the easy-magnetization direction {[}111{]} curve
near $\unit[300]{mT}$ as seen in Figure \ref{fig:Field-dependence-of}.
This anisotropy field of $\mathrm{U_{4}Ru_{7}Ge_{6}}$ is an unprecedentedly
low value ever observed for a uranium intermetallic compound. Moreover,
we observed an entirely isotropic paramagnetic susceptibility and
electrical resistivity which up to our best knowledge has not been
reported for any uranium 5$f$-electron ferromagnet.

One may argue that the almost isotropic magnetism in $\mathrm{U_{4}Ru_{7}Ge_{6}}$
is a consequence of the itinerant character of U 5$f$-electron magnetic
moment. The spontaneous magnetic moment of this compound is $\sim\unit[0.85]{\mu_{\mathrm{B}}/f.u.}$,
which provides an average moment of $\sim\unit[0.21]{\mu_{\mathrm{B}}/U}$
ion when supposing negligible contributions from Ru and Ge ions. This
value is 4 times larger than the U-born magnetic moment of the itinerant
5$f$ electron ferromagnet$\mathrm{UNi_{2}}$\cite{RN41,RN42,RN43,RN44}
which, in contrary, exhibits huge magnetocrystalline anisotropy with
the anisotropy field $\gg\unit[35]{T}$\cite{RN44}.

Both our magnetization data and ab initio calculations clearly show
that the easy magnetization direction of $\mathrm{U_{4}Ru_{7}Ge_{6}}$
in the ground state is {[}111{]}. As a result of magnetoelastic interaction\cite{RN60},
the ferromagnetic ordering at temperatures below $T_{\mathrm{C}}$
is accompanied by the spontaneous magnetostriction causing a distortion
of a crystal lattice related to the easy-magnetization direction.
These distortions are in ferromagnetic materials possessing cubic
crystal structure in paramagnetic state (at temperatures $T>T_{\mathrm{C}}$)
the spontaneous magnetostriction leads to a distortion lowering the
paramagnetic cubic to tetragonal, orthorhombic and rhombohedral symmetry
for the {[}001{]}-, {[}110{]}- and {[}111{]}-easy-magnetization direction,
respectively. The expected rhombohedral distortion of the cubic $\mathrm{U_{4}Ru_{7}Ge_{6}}$
lattice in the ferromagnetic ground state is so tiny that it falls
within the experimental error of a standard X-ray diffraction but
is clearly indicated by thermal expansion results at low temperatures.
As a consequence of the distortion the one equivalent crystallographic
site common for all U ions in the cubic lattice splits into two inequivalent
ones which is confirmed by ab initio calculations.

Our magnetization results also reveal that the easy-magnetization
direction holds onto the {[}111{]} axis only at low temperatures up
to $T_{\mathrm{r}}$ ($=\unit[5.9]{K}$) whereas at higher temperatures
up to $T_{\mathrm{C}}$ the easy-magnetization direction is unambiguously
along {[}001{]} and the paramagnetic cubic lattice is tetragonally
distorted along this direction. This finding is in good agreement
with the theoretical calculations which reveal the excited state with
the {[}001{]} easy magnetization $\unit[0.9]{meV}$ above the ground
state.

The magnetic moment reorientation transition in $\mathrm{U_{4}Ru_{7}Ge_{6}}$
at $T_{\mathrm{r}}$ is manifested in specific features (anomalies)
which we observed in the temperature dependencies of magnetization
(Figure \ref{fig:Temperature-dependence-of}a), AC susceptibility
(Figure \ref{fig:Temperature-dependence-of}b), specific heat (Figure
\ref{fig:Temperature-dependence-of-2}), thermal expansion (Figure
\ref{fig:Linear-thermal-expansion}) and electrical resistivity (Figure
\ref{fig:Temperature-dependence-of-3}). It is in fact an order-to-order
magnetic phase transition accompanied by structural distortion due
to notable magnetoelastic coupling. These phase transitions are known
to be of the first order type (e.g. $\mathrm{HoAl_{2}}$\cite{RN61}).
The thermal expansion anomalies seen in Figure \ref{fig:Linear-thermal-expansion}
at $T_{\mathrm{C}}$ and $T_{\mathrm{r}}$, respectively, may be viewed
as an illustrative examples of the second and first order type phase
transitions. However, the first order phase transition is to be accompanied
by latent heat which we were unable to detect by detailed specific-heat
measurement and the $T_{\mathrm{r}}$ related specific-heat anomaly
is considerably broader than expected for a first order phase transition.
We attribute the lack of observables pointing to the presence of latent
heat to the complex domain structure processes during the spin reorientation
in the multidomain sample in the vicinity of $T_{\mathrm{r}}$ and
our tentative determination has to be confirmed by a designed method
allowing determination of order type by other means (e.g. phase coexistence
in $\mu\mathrm{SR}$).

The anisotropy field values in U ferromagnets are typically hundreds
T whereas in $\mathrm{U_{4}Ru_{7}Ge_{6}}$ it is roughly 3 orders
of magnitude smaller value. When inspecting crystal structures we
observe that in all cases of the U ferromagnets characterized by high
values of anisotropy field the U ions have some U nearest neighbors.
Contrary, the individual U ions in $\mathrm{U_{4}Ru_{7}Ge_{6}}$are
buried inside the Ru and Ge polyhedra preventing direct connection
to any nearest U ion which should have consequences for magnetism\cite{RN2}.

The direct 5$f$-5$f$ overlap of U electron orbitals is probably
behind the huge magnetic anisotropy of U compounds. The symmetry of
the network of U nearest neighbors determines the type of magnetic
anisotropy in these materials\cite{RN20}. The driving mechanism of
magnetic anisotropy in $\mathrm{U_{4}Ru_{7}Ge_{6}}$ is presumably
the hybridization of U 5$f$-electron states with the 4$d$-electron
states of surrounding Ru ions which are in hexagonal arrangement,
perpendicular to the easy magnetization direction {[}111{]}.

Onset of itinerant electron ferromagnetism is usually accompanied
by a positive spontaneous magnetovolume effect\cite{RN62,RN63}. Also
our thermal expansion data show this tendency despite the negative
value of $\lambda_{\mathrm{S\left[001\right]}}$. Unfortunately, the
measurements using dilatometer cannot be extended to temperatures
lower than $T_{\mathrm{r}}$ because the body diagonals representing
the {[}111{]} easy magnetization direction are not perpendicular and
therefore an experiment analogous to that for $T_{\mathrm{r}}<T<T_{\mathrm{C}}$
is not accessible.

\section{Conclusions}

We have demonstrated by experiment and by calculations the almost
isotropic ferromagnetism in $\mathrm{U_{4}Ru_{7}Ge_{6}}$ which represents
a case contrasting the huge magnetocrystalline anisotropy in so far
reported ferromagnetic uranium compounds exhibiting by rule at least
3 orders of magnitude larger values of anisotropy field. The ground-state
easy-magnetization direction is along the {[}111{]} axis of the cubic
lattice as found both by experiment and by calculations. At $T_{\mathrm{r}}=\unit[5.9]{K}$
the easy magnetization direction changes for {[}100{]} which holds
at temperatures up to $T_{\mathrm{C}}$ as found by experiments. This
is in agreement with calculations revealing that the moment pointing
to the {[}001{]} direction is characteristic for the excited state
which is only $\unit[0.9]{meV}$ above the ground state. This spin
reorientation transition is significantly projected in low-field magnetization,
AC susceptibility and thermal-expansion data and causes also a weak
anomaly at $T_{\mathrm{r}}$ visible in the temperature dependence
of specific heat and electrical resistivity, respectively.

The magnetoelastic interaction induces a rhombohedral (tetragonal)
distortion of the paramagnetic cubic crystal lattice in case of the
{[}111{]}({[}001{]}) easy-magnetization direction. The rhombohedral
distortion leads to emergence of two crystallographically inequivalent
U sites which is also in agreement with results of calculations. We
propose a scenario of the very weak magnetic anisotropy of $\mathrm{U_{4}Ru_{7}Ge_{6}}$
connected with the lack of direct 5$f$-5$f$ overlaps of orbitals
of nearest U \textendash{} U neighbors. These are prevented by the
specific crystal structure in which the individual U ions are trapped
inside of the Ru, Ge polyhedra. 
\begin{acknowledgments}
The authors are indebted to Jan Prokle\v{s}ka for checking the manuscript,
providing critical insight and constructive suggestions. This work
was supported by the Czech Science Foundation Grant No. P204/16/06422S
and by Charles University, project GA UK No. 720214. Experiments were
performed in MLTL (\url{http://mltl.eu/}) which is supported within
the program of Czech Research Infrastructures (Project No. LM2011025).
\end{acknowledgments}

\bibliographystyle{apsrev4-1}

\end{document}